\documentstyle[12pt,epsf, multicol]{article}
\begin{document}
\newcommand{\rf}{\par\noindent\hangindent 15pt {}}
\newcommand{\rl}{\par\noindent\hangindent 50pt {}}

\thispagestyle{empty}

\begin{center}
{\large 
{\bf The Galactic Center:  A Laboratory for Fundamental Astrophysics and 
Galactic Nuclei} \\ 
{\it  An Astro2010 Science White Paper} \\
}
\end{center}
\bigskip
\noindent
{\centering \leavevmode
\hspace*{.25\columnwidth}
\epsfxsize=.5\columnwidth \epsfbox{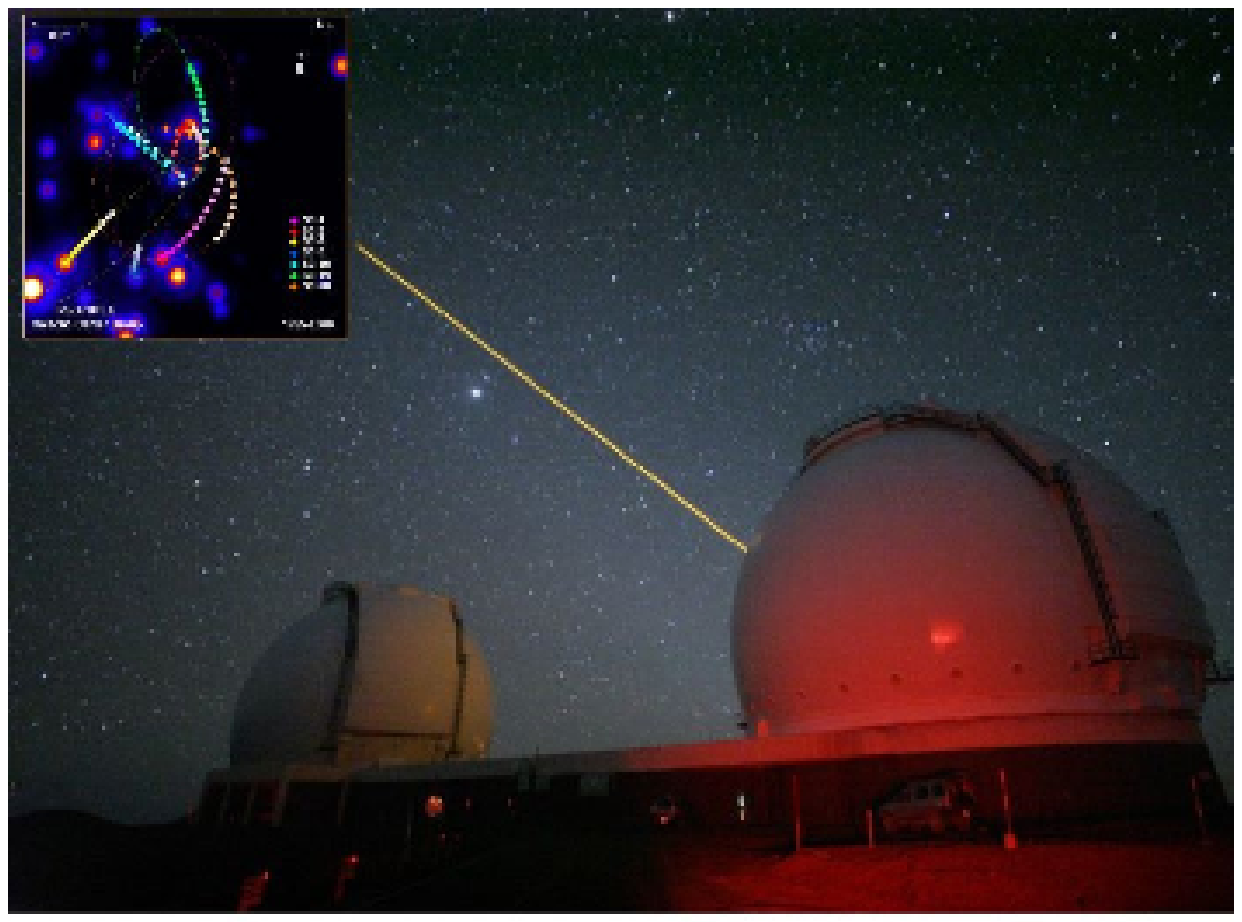}
}
\begin{center}
\underline {Authors:} 

\noindent
Andrea Ghez (UCLA; ghez@astro.ucla.edu),
Mark Morris (UCLA), 
Jessica Lu (Caltech), 
Nevin Weinberg (UCB),
Keith Matthews (Caltech),
Tal Alexander (Weizmann Inst.),
Phil Armitage (U. of Colorado),
Eric Becklin (Ames/UCLA),
Warren Brown (CfA),
Randy Campbell (Keck) 
Tuan Do (UCLA),
Andrea Eckart (U. of Cologne),
Reinhard Genzel (MPE/UCB),
Andy Gould (Ohio State),
Brad Hansen (UCLA),
Luis Ho (Carnegie),
Fred Lo (NRAO),
Avi Loeb (Harvard),
Fulvio Melia (U. of Arizona),
David Merritt (RIT),
Milos Milosavljevic (U. of Texas), 
Hagai Perets (Weizmann Inst.),
Fred Rasio (Northwestern),
Mark Reid (CfA),
Samir Salim (NOAO),
Rainer Sch\"odel (IAA), 
Sylvana Yelda (UCLA)

\bigskip
\underline {Submitted to Science Frontier Panels on} \\
(1) The Galactic Neighborhood (GAN) \\
(2) Cosmology and Fundamental Physics (CFP)

\bigskip
\underline {Supplemental Animations:} \\
{\it http://www.astro.ucla.edu/$\sim$ghezgroup/gc/pictures/Future\_GCorbits.shtml}
\end{center}

\pagebreak

\setcounter{page}{1}

\begin{center}
{\large{\bf Abstract}}
\end{center}
As the closest example of a galactic nucleus, the Galactic center presents
an exquisite laboratory for learning about supermassive 
black holes (SMBH) and their environs.  In this document, we describe how
detailed studies of stellar dynamics deep in the potential well of a galaxy
offer several exciting directions in the coming decade.
First, it will be possible to obtain precision measurements of the Galaxy's
central potential, providing both a unique test of General Relativity (GR)
and a detection of the extended dark matter
distribution that is predicted to exist around the SMBH. 
Tests of gravity have not previously been possible on
scales larger than our solar system, or in regimes where the gravitational
energy of a body is $> \sim$1\% of its rest mass
energy.  Similarly, only upper limits on the extended matter distribution
on small scales currently exist; detection of dark matter on these 
scales is an important test of Lambda-CDM and
the detection of stellar remnants would reveal a population that may
dominate the stellar dynamics on the smallest scales.
Second, our detailed view of the SMBH and its local gas and stellar environment
provides insight into how SMBHs
at the centers of galaxies form, grow and interact with their environs 
as well as on the exotic processes at work in the densest stellar clusters
in the Universe.
The key questions, still unanswered, of when and how SMBHs formed in the early
universe, and the myriad ways in which feedback from SMBHs can affect
structure formation, can be informed by directly observing the physical 
processes operating at the SMBH.
The keys to realizing these science objectives are
(1) observing modes and program support on telescopes that will allow 
for long time-baseline observations and careful monitoring
of orbits during closest approach, 
(2) larger field of view (FOV) Adaptive Optics fed Integral Field Units,
(3) improved strehl
ratio (SR) performance for adaptive optics facilities on existing large 
ground-based telescopes, 
(4) more realistic multi-scale and multi-physics numerical
modeling of the stellar dynamics, stellar evolution, and star formation taking place in the dense star cluster
around a SMBH,
and
(5) higher angular resolution from high sensitivity interferometry,
and, most importantly, a future extremely large (30-m) ground-based telescope
(ELT), which is critical to several opportunities outlined herein.


\bigskip
\noindent
{\large {\bf 1 $~ ~ $Introduction}} 

The proximity of our Galaxy's center presents a unique opportunity to study a
galactic nucleus with orders of magnitude higher spatial resolution
than can be brought to bear on any other galaxy.  
Over the past decade tremendous advances in high angular resolution infrared 
astronomy have led to further improvements in the spatial 
detail by
yet another order of magnitude (see Figure 1 \& 2).
The  advantage of the milli-parsec scales that can be probed 
has led to a number of exciting results, 
including

\noindent
{\it [1]} an extremely strong case for the existence of a supermassive black hole 
(SMBH) at the center of a normal galaxy (without an AGN), 
based on measurements of stellar orbits
(Ghez et al. 2000, 2003a, 2005a, 2008; Eckart et al. 2002;
Sch{\"o}del et al. 2002, 2003; Eisenhauer et al. 2003, 2005;
Gillessen et al. 2009);



\noindent
{\it [2]} the first infrared detection of SgrA*,
the radio source associated with the central SMBH,
providing among the best constraints on theoretical models
for low accretion rate flows - important for galactic
nuclei and also for X-ray binaries; also detection of
its dramatic short-timescale infrared variations 
 showing red noise behavior and a power-law break consistent with the X-ray
 variability seen in AGNs (Genzel et al.\ 2003; Ghez et al.\ 2004; 
Eckart et al.\ 2006; Yusef-Zadeh et al.\ 2006; Hornstein et al.\ 2007; Do et al.\ 2009; 
Meyer et al.\ 2008, 2009); 

\noindent
{\it [3]} discovery that the stars within 0.01 pc
(0.$\tt''$25) of the central SMBH are young
stars (B stars; Ghez et al. 2003a; Eisenhauer et al. 2005;
Gillessen et al. 2008; Do et al. 2008) and may be the remainders of the
dynamical process that leads to the ejection of the hypervelocity stars 
that have recently been discovered in the Galactic halo (Brown et al. 2005, 
2007, 2009; Hirsch et al. 2005; Edelmann et al. 2005)

\noindent
{\it [4]} discovery of a disk of massive young stars orbiting the SMBH at radii 
between $\sim$ 0.04 pc (1$\tt''$) and 0.5 pc,
 likely indicating an {\it in situ} formation event 5 - 7 $\times~10^6$
 years ago (Levin \& Beloborodov 2003; Genzel et al. 2003b; 
Paumard et al. 2006; Lu et al. 2009);


\noindent
The Milky Way has therefore become a laboratory for learning by
example about SMBHs and their environs at the centers of other galaxies.
With both existing and upcoming facilities, the next decade 
of high angular resolution infrared observations offers a 
promising new era of discovery in Galactic center research.

\bigskip

\noindent
{\centering \leavevmode
\parbox[b]{0.45\columnwidth}{
{\bf Figure 1:} A composite near-infrared (HKL') image of the central 
10$\tt''$$\times$10$\tt''$ (0.4 pc $\times$ 0.4 pc) of our Galaxy obtained with the laser 
guide star adaptive optics system at the W. M. Keck telescope.
With this and similar set-ups, it has been possible to demonstrate the 
existence of a supermassive black hole, to detect the infrared emission 
associated 
with the black hole, and begin to study the paradoxical young stars in this 
region.  However, even at this resolution (fwhm $\sim$ 60 mas), 
source confusion is the dominant source of positional uncertainty.  
}
\hspace*{.05\columnwidth}
\epsfxsize=0.475\columnwidth \epsfbox{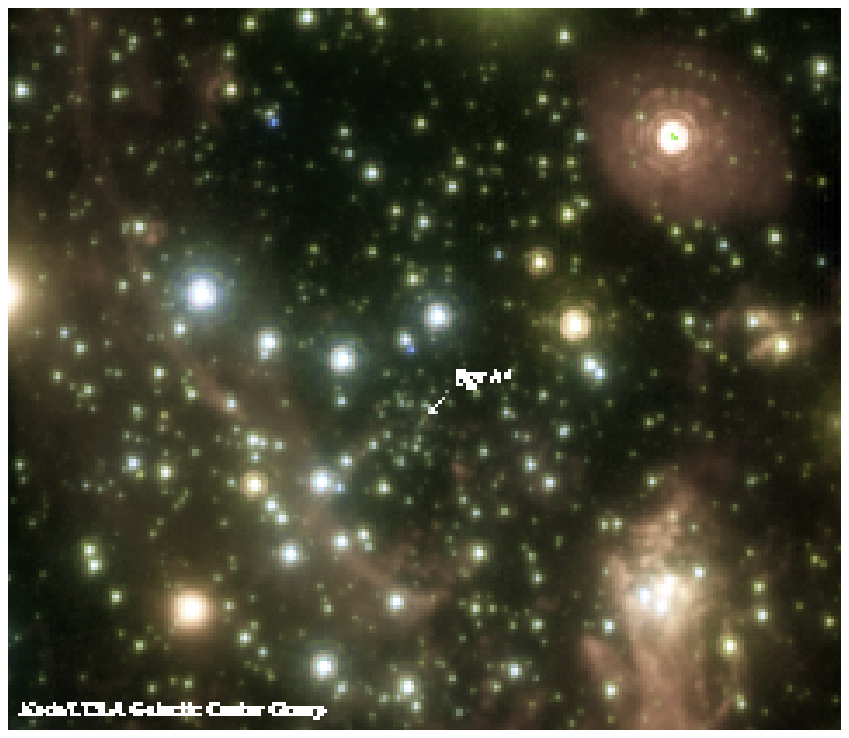}
}

%
%
%
%

\bigskip
\noindent
{\large {\bf 2 $~ ~$Tests of Relativity \& the Dark Matter in the Galactic Center}}

Ever since the first measurements of proper motions of stars within
0.$\tt''$3 (2400 AU) of the center of our Galaxy (Eckart \& Genzel 1997;
Ghez et al. 1998), the prospect of using stellar orbits to make ultra-precise
measurements of the distance to the Galactic center (R$_o$),
and, more ambitiously, to measure
post-Newtonian effects 
has been anticipated
by many under the assumption that radial velocities and more accurate astrometry
would eventually be obtained
(Jaroszynski 1998, 1999; Salim \& Gould 1999; Fragile \& Mathews 2000;
Rubilar \& Eckart 2001; Weinberg et al.\ 2005; Zucker
\& Alexander 2006; Kraniotis 2007; Nucita et al.\ 2007; Will 2008).

\underline{General Relativity}:
Of the theories describing the four fundamental forces of nature, the theory
that describes gravity, general relativity (GR), is the least tested.
In particular, GR has not been tested on the mass scale of an SMBH.
Recent results suggest that, within the next decade, it may be possible 
to begin to test GR on this scale with measurements of stars and gas in the 
immediate vicinity of the SMBH at the center of the Milky Way.
The highly eccentric 15 yr orbit of the star S0-2 brings it within 100 AU
of the SMBH, corresponding to $\sim$1000 times the black hole's
Schwarzschild radius (i.e., its event horizon).   Studying the pericenter
passage of S0-2 and other high-eccentricity stars therefore offers
an opportunity to test GR in a unique regime.
With a decade of
$\sim$100 $\mu$as astrometric and $\sim$20 km/sec RV precision, one can
detect the astrometric signal of prograde GR precession
(see, e.g., Weinberg et al.\ 2005) and the influence on RV measurements
from the special relativistic transverse Doppler
shift and the general relativistic gravitational redshift
(Zucker \& Alexander 2006).

\underline{Extended Dark Matter Distribution}:
%
%
The observed stellar sources around the SMBH may represent only
a fraction of the total matter content. Within 0.01 pc of the SMBH,
 the matter density may instead be dominated either by massive compact
remnants (5-10 M$_\odot$ black holes) or cold dark matter. Theoretical estimates
suggest that  $\sim$1000 M$_\odot$ of compact remnants and as much as $\sim$1000 M$_\odot$ of cold dark
matter particles may  reside in the inner 0.01 pc of the Galactic center
(Morris 1993; Miralda-Escud\'e \& Gould 2000;
Gondolo \& Silk 1999; Ullio et al.\ 2001; Merritt et al.\ 2002;
Gnedin \& Primack 2004; Bertone \& Merritt 2005).  
This extended matter around the central
black hole has not yet been detected (upper limits in Mouawad et al.\ 2005;
Ghez et al.\ 2008; Gillessen et al.\ 2009).
Detection of the dark matter is an important test of Lambda-CDM and
the detection of stellar remnants would reveal a population that may
dominate the stellar dyanmics on the smallest scales; these two components
can be disentangled through the expected scattering events from the latter
population (Weinberg et al. 2005).
If the concentration of extended matter at the Galactic center matches
theoretical predictions, its influence on the orbits
(retrograde precession) will be detectable, even in the presence of GR effects,
with a decade of
$\sim$100 $\mu$as astrometric and $\sim$20 km/sec RV precision orbital
motion measurements.  

\underline{R$_o$ and Galactic Structure}:
Precision measurement of R$_o$ is important as it affects almost all questions 
of Galactic structure and mass. For instance, a 1\% measurement
would enable one to determine
the size and shape of the Milky Way's several 100 kpc-scale dark matter halo
(see, e.g., Olling \& Merrifield 2000).
The halo shape tells us about the nature of
dark matter (e.g., the extent to which it self-interacts) and the process of
galaxy formation (how the dark matter halo relaxes following mergers).
Currently the shape is very poorly constrained.
Furthermore, a precision measurement of R$_0$, in combination with the proper
motion of SgrA* relative to background quasars (e.g., Reid \& Brunthaler 2004), leads directly to a
precision measurement of the Galaxy's local rotation speed.

\underline{Requirements to Achieve the Scientific Goal}:
The use of Adaptive Optics (AO) with large ground-based telescopes should
allow many of these precision measurements of the central potential of our 
Galaxy for the first time within the next decade (see Figure 2).  
While current AO systems can deliver the required astrometric 
and radial velocity precision,
the accuracy of the astrometric measurements are degraded to $\sim$0.5 mas
(for S0-2, which is the brightest of the short period stars) or more due to 
source confusion in this high density region (Ghez et al.\ 2008; 
Gillissen et al.\ 2009).   Future AO systems that deliver
higher strehl ratios on existing ground-based telescopes, such as NGAO 
on Keck, will be able to overcome this source of confusion for S0-2.
Future extremely large ground-based telescope (ELT), such as TMT/GSMT, will 
provide the required accuracy for the larger population of fainter
short-period stars, which are necessary to provide a complete probe
of all the possible post-Newtonian effects 
measurable by the orbits
of known stars
(special relativistic transverse Doppler
shift, the general relativistic gravitational redshift, the general
relativistic prograde motion
of periapse, and the retrograde motion of the periapse from the extended
mass distribution).  They will also allow one to determine orbits closer
in to the black hole, where GR effects will be most prominent.
Interferometry could also provide additional
resolution necessary to disentangle source confusion for the brighter
sources at the critical
periapse passsage phase of the orbit.  However the FOV and 
angular resolution of an ELT will be criticial for establishing the
astrometric reference frame in this crowded and dynamic field.

\noindent
{\centering \leavevmode
\epsfxsize=0.33\columnwidth \epsfbox{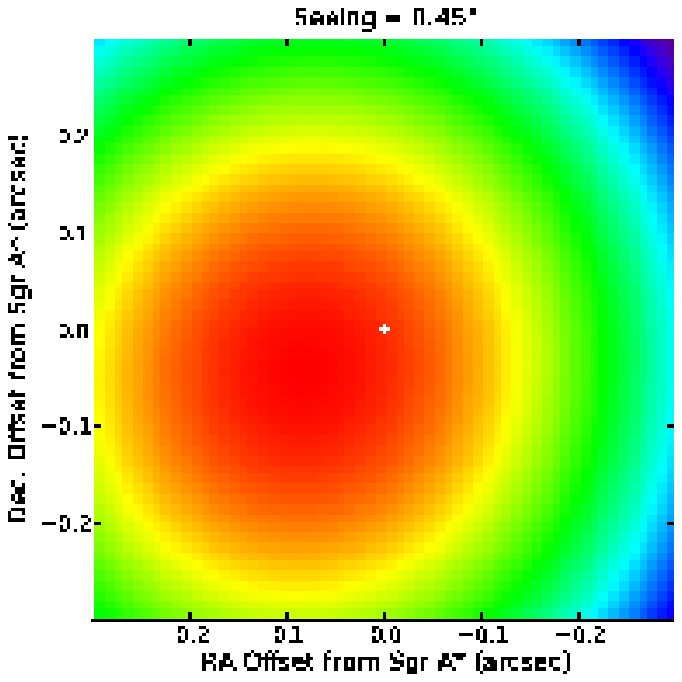} 
\epsfxsize=0.33\columnwidth \epsfbox{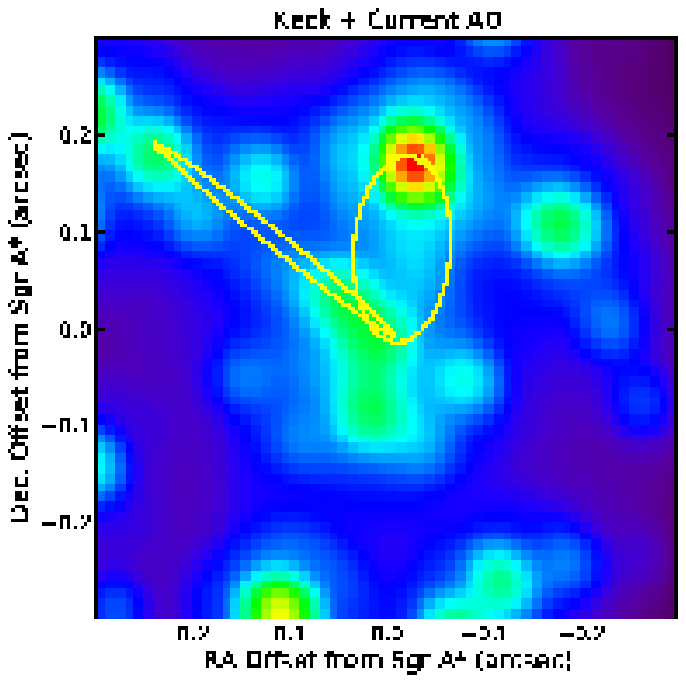}
\epsfxsize=0.33\columnwidth \epsfbox{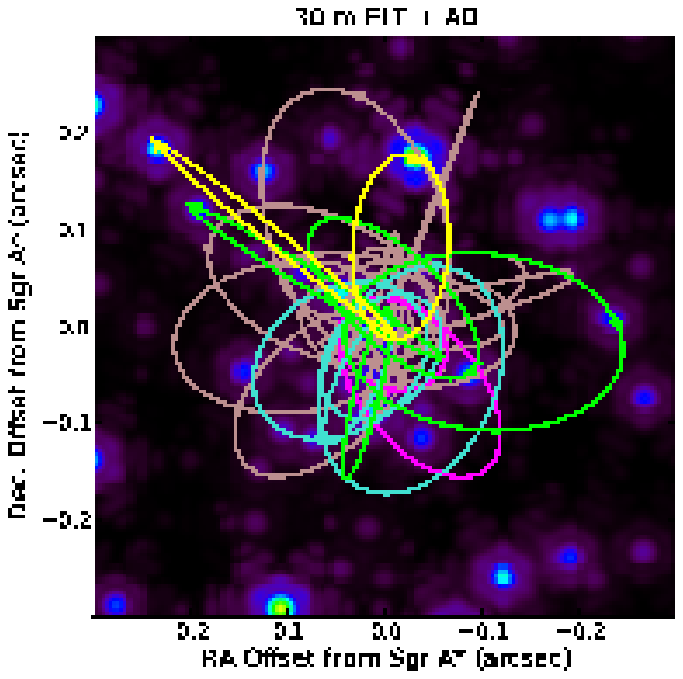}
}

\noindent
{\bf Figure 2:} A 0.$\tt''$6$\times$0.$\tt''$6 (0.02 $\times$ 0.02 pc)
region centered on the Milky Way's SMBH as imaged at 2.1 $\mu$m 
under seeing-limited
conditions ($\theta_{fwhm} \sim$0.$\tt''$45), with the current Keck+AO system
($\theta_{fwhm} \sim$0.$\tt''$06 and SR = 0.3), and with a 
future 30-m telescope with a multiple laser AO system ($\theta_{fwhm} \sim$0.$\tt$015 and SR = 0.8).  Overlaid are all 
known or examples of expected orbits with orbital periods less than 23 years
that are expected to be detectable in each image 
(14$<$K$<$16, yellow;
K$<$17, green; 
K$<$18, cyan; K$<$19, magenta; K$<$20, tan).
A future 30-m telescope would not only increase 
the number of measurable short period orbits by an order of magnitude, 
but should also find systems that orbit
the SMBH much deeper in the central potential (with orbital periods that are a 
factor of 5 smaller).   These systems are particularly helpful for measurements 
of post-Newtonian effects (GR and extended mass distribution).
(Animated version and stills without overlays: 
{\it http://www.astro.ucla.edu/$\sim$ghezgroup/gc/pictures/Future\_GCorbits.shtml})


\bigskip
\noindent
{\large {\bf 3 $~ ~$The Galactic center as a Case Study of Interactions 
Between the SMBH and its Environs}}

The center of the Milky Way offers a unique possibility to directly
study the structure and dynamics of stars surrounding a SMBH.
With Adaptive Optics, it is now possible to measure the three-dimensional
kinematics and spectral class of stars that reside in key structures of 
interest, including, for the dynamically relaxed stellar population, 
the central stellar cusp (see \S3.1), and, for the young, dynamically unrelaxed 
stars, the stellar disks and associated structures (see \S3.2).  
Given the unprecedentedly detailed measurements of stellar orbits in a 
cluster, the Galactic center is also an ideal test-bed for 
multi-particle dynamics.


\underline{A Unique View of a Stellar Cusp Around a Central Black Hole}

Theoretical work suggests that a stellar cusp is a distinct signature of a
central SMBH embedded in a stellar cluster
and that the properties of this cusp, particularly the index of the
power-law that describes the stellar density,
should reflect the formation history of the system
(e.g., Young 1980; Lee \& Goodman 1989; Quinlan et al.\ 1995; Bahcall \& Wolf
1976, 1977).  If these theoretical predictions are confirmed for a
relaxed stellar population at the Galactic center, this would provide a 
means to calibrate a method for determining the presence of massive black 
holes in other stellar systems, for example.

Efforts thus far to characterize the central stellar cusp at high 
angular resolution from number counts suggest that the radial profile of old 
stars is significantly flatter than that expected in a collisionally 
relaxed population around a SMBH 
(Genzel et al.\ 2003; Sch\"odel et al.\ 2007, 2009; 
Do et al.\ 2009).  Five possibilities exist for explaining this fact:
(1) mass segregation with stellar remnants such as black holes or neutron stars,
(2) infall of an intermediate-mass black hole, (3) collision of these 
evolved giants with stars or binaries in this high-density region, (4) tidal stripping of
stars with high eccentricities that bring them to the tidal radius of the black hole and (5) the most recent merger event that created the bulge occurred 
too recently for a cusp to re-form after being destroyed by a binary SMBH
(e.g. Murphy et al.\ 1991; Alexander 2005; Alexander \& Hopman 2008).
To differentiate between the collisional and mass-segregation scenarios,
it is necessary to obtain more 
complete spectral information, such that the sizes and masses of the late-type
stars can be determined, and to obtain more complete kinematic information,
such that the degeneracies in de-projecting two-dimensional number counts 
into a three-dimensional stellar distribution can be broken. 
These measurements will give us an unprecedented view of a stellar cusp around
a SMBH.

\underline{Star Formation in the Vicinity of a Supermassive Black Hole}

Within the strong tidal field of our Galaxy's central SMBH there are two,
possibly related, populations of young massive stars, whose presence 
is a puzzle.
The bright post-main-sequence stars (Wolf Rayet, OB giants and supergiants),
residing between 1 and 14$\tt''$ (0.04 - 0.5 pc) and
having ages of $\sim$6 Myr,
were the first to be recognized
(e.g., Allen et al.\ 1990; Krabbe et al.\ 1995;
Blum et al.\ 1995; Tamblyn et al.\ 1996;
Najarro et al.\ 1997; Paumard et al. 2006; Tanner et al.\ 2006).
More recently, a fainter population of main-sequence B stars has been
discovered in the inner 1$\tt''$ (0.04 pc) around SgrA*
(referred to as the SgrA* stellar cluster; Ghez et al.\ 2003;
Eisenhauer et al.\ 2005;
Gillessen et al.\ 2008).
The origin of these two populations of young stars
is difficult to explain since
the gas densities observed today are orders of magnitude too low
for a gas clump to overcome the extreme tidal forces and
collapse to form stars
(e.g. Sanders 1992; Morris 1993; Genzel et al.\ 2003; Ghez et al.\ 2005;
Alexander 2005).
Populations of young stars have also been observed in the nuclei of other
galaxies, such as in M31 (Bender et al.\ 2005), suggesting that star formation
near a SMBH may be a common, but not well-understood,
phenomenon in galaxy evolution.

Stellar kinematics provide a powerful probe of the star formation process
and have revealed that the majority of
young stars outside the central 1$\tt''$ orbit the SMBH
in one, and possibly two, roughly perpendicular,
stellar disks (Levin \& Beloborodov 2003; Genzel 2003; Paumard et al.\
2006; Lu et al. 2009; Bartko et al.\ 2009).
These observations have inspired a wide variety of star formation theories 
including formation within a pre-existing massive, 
self-gravitating gas disk, a radially infalling cloud, 
or an inspiralling cluster
(e.g., Sanders et al.\ 1998; Gerhard 2001; Kim \& Morris 2003;
Levin \& Beloborodov 2003; 
Portegies Zwart et al.\ 2002; McMillan \& Portegies Zwart
2003; Hansen \& Milosavljevic 2003; G\"urkan \& Rasio 2005;
Nayaksin \& Cuadra 2005; Berukoff \& Hansen 2006; Yu et al.\ 2007; 
Levin 2007 Alexander et al.\ 2008;
Cuadra et al.\ 2008; Bonnell \& Rice 2008;
Yusef-Zadeh et al.\ 2008; Perets et al.\ 2008b).
While kinematic structure of the young star population, 
in principle, can distinguish
between these models, there is tremendous debate over the 
detailed properties of these stars.  For example, it is not
currently clear whether there is 
indeed a second stellar disk of young stars, and similarly,
whether there is a warp in the first, 
more compact disk at larger radii, due to the incomplete census of young stars 
outside the central 0.1 pc ($\sim$3 arcsec) and the more limited kinematic
information at these larger radii.

The origin of the young stars within the central 1$\tt''$, the SgrA* cluster,
has proven to be even more difficult to understand.
A wide range of models
have been proposed to explain them, including
(1) collisional mergers of multiple low-mass stars
to form a higher-mass hot star akin to a ``blue straggler" (Genzel et al.\ 2003),
(2) migration through a parent disk via type-I or runaway migration and subsequent Rauch-Tremaine resonant relaxation (Levin 2007, Hopman \& Alexander 2006),
(3) precession and Kozai interactions (L\"ockmann et al.\ 2008 [but see Chang 2008]),
(4) three-body
exchanges between binary stars and the SMBH or stars and a BH binary
(e.g., Gould \& Quillen 2003; 2003; Perets et al.\ 2008a; Merritt et al.\ 2009).
Models that include three-body exchanges
lead to a natural connection between the B stars
observed at the
Galactic center and those detected as hyper-velocity stars in
the halo of our Galaxy (e.g., Yu \& Tremaine 2003;
Brown et al.\ 2005, 2007, 2008; Hirsch et al.\ 2005; Edelmann et al.\ 2005;
Perets et al.\ 2007; Ginsburg \& Loeb 2007;
Blecha \& Loeb 2008; Sesana et al.\ 2007, 2008; see also
O'Leary \& Loeb 2008).
With full orbital solutions for only $\sim$15 known members of the
SgrA* cluster,
it is difficult to understand their statistical
properties (Ghez et al.\ 2005; Eisenhauer et al.\ 2005; Gillessen et al.\ 2008).
Future observations with improved AO on existing telescopes should
double the number of orbital estimates for cluster members and a
future even larger ground-based telescope should increase this number
by an order of magnitude, thereby allowing mass-dependent effects to be
detected.


\underline{Requirements to Achieve the Scientific Goal}:
The two key developments that would benefit the goals in this section are
(1) larger format AO-fed IFUs and (2) higher precision astrometric
measurements with improved AO and a larger ground-based telescope.

With spatial scales of $\sim$1 pc (20 arcsec) and high stellar densities, the 
key
structures at the Galactic center discussed above require AO-fed integral field units (IFU) to cover them spectroscopically.  However, with fields of $\sim$1-2 arcsec, current AO-fed IFU have
only begun to probe these structures.  Over the coming decade, complete
spectroscopic coverage will be a high priority to obtain the spectral types
and radial velocities of stars in the central nuclear cluster.  Key 
developments that benefit this work would be larger FOV AO-fed IFUs on
existing large ground-based telescopes; this would take
advantage of the AO-corrected FOV for existing AO systems, which are large
compared to current AO-fed IFUs.  

Astrometric measurements offer valuable insight into the three dimensional
distribution of stars.  Our Galaxy is the only galactic nucleus in which
the three dimensional velocities of the individual 
stars can be probed.  While this is a tremendous advantage over other
galaxies, it is possible to obtain an even more direct constraint on
the stellar kinematics by measuring individual stellar
orbits.  Complete stellar orbits are now possible out to 
a radius of $\sim$0.04 pc (1 arcsec) and will be pushed out to larger radii with
increased time baseline and precision of the measurements.  Therefore
key developments that benefit this work are those that improve
the astrometric precision, such as higher strehl ratio AO
systems on existing large ground-based telescopes and an 
even larger ground-based
telescope with AO (since source confusion is a dominant source of uncertainty
over the entire region of interest; see Ghez et al.\ 2008; Gillessen et al.\
2009; Sch\"odel et al.\ 2009).  

\bigskip
\begin{multicols}{2}
\rf{Alexander, T. 2005, Physics Reports, 419, 65}
\rf{Alexander, T., Hopman, C., 2008, in press}
\rf{Alexander, R. et al.\ 2008, ApJ, 674, 927}
\rf{Bonnell \& Rice  2008, Science, 321, 1060}
\rf{Bahcall, J.\ \& Wolf, R.\ 1977, ApJ, 216, 883}
\rf{Bender, R. et al.  2005,
ApJ, 631, 280}
\rf{Bertone \& Merritt 2005, PRD, 72, 103502}
\rf{Berukoff \& Hansen 2006, ApJ, 650, 901}
\rf{Blecha, L., Loeb, A. 2008, MNRAS, 390, 1311}
\rf{Blum, R. et al. 1995, ApJ, 441, 603}
\rf{Brown, W. et al. 2005, ApJ, 622, L33;
2007, ApJ, 671, 1708;
2009, ApJ, 690, L69}
\rf{Chang, P. 2008, in press, arXiv:0811.0829}
\rf{Cuadra, J. et al. 2008, MNRAS, 388, L64}
\rf{Do, T. et al.  2009, ApJ, 691, 1021}
\rf{Eckart et al.\ 1997, MNRAS, 284, 576; 2002, MNRAS, 331, 917;
2006, A\&A, 450, 535}
\rf{Edelmann, H. et al. 2005, ApJ, 634, L181}
\rf{Eisenhauer, F. et al.
2003, ApJ, 597, L121;
2005, ApJ, 628, 246-259}
\rf{Fragile \& Mathews 2000, ApJ, 542, 328}
\rf{Genzel, R., et al.\
2003a, Nature, 425, 859;
2003b, ApJ, 594, 812}
\rf{Gerhard, O. 2001, ApJ, 546, L39}
\rf{Ghez, A.\ et al.\ 1998, ApJ, 509, 678; 
2000, Nature, 407, 349;
2003, ApJ, 586, L127;
2004, ApJ, 601, L159;
2005a, ApJ, 620, 744;
2005b, ApJ, 635, 1087;
2008, ApJ, 689, 1044}
\rf{Gillessen, S. et al. 2009, in press}
\rf{Ginsburg \& Loeb  2007, MNRAS, 376, 492}
\rf{Gnedin, O. Y., Primack, J. R. 2004, PRL, 93, 061302} 
\rf{Gondolo, P.~\& Silk, J.\ 1999, Physical Review Letters, 83, 1719}
\rf{Gould, A. \& Quillen, A., 2003, ApJ, 592, 935}
\rf{G{\" u}rkan \& Rasio 2005, ApJ, 628, 236}
\rf{Hansen \& Milosavljevi{\' c} 2003, ApJ, 593, L77}
\rf{Hirsch, H. et al. 2005, A\&A, 444, L61}
\rf{Hopman \& Alexander 2006, ApJ, 645, 1152}
\rf{Hornstein, S. et al.\ 2007, ApJ, 667, 900}
\rf{Jaroszynski, M., 1998, ACA, 48, 653; 1999, ApJ, 521, 591}
\rf{Kim, S.~S., Morris, M.\ 2003
ApJ, 597, 312}
\rf{Krabbe, A. et al. 1995, ApJ, 447, L95}
\rf{Kraniotis  2007, Classical Quantum Gravity, 24, 1775}
\rf{Lee, M. \& Goodman, J. 1989, ApJ, 343, 594}
\rf{Levin \& Beloborodov 2003, ApJ, 590, L33}
\rf{Levin, Y. 2007, MNRAS, 374, 515}
\rf{L{\" o}ckmann, U. et al. 2008, ApJ, 683, L151}
\rf{Lu, J. et al.  2009, ApJ, 690, 1463}
\rf{Marrone, D. et al. 2008, ApJ, 682, 373}
\rf{McMillan \& Portegies Zwart 2003, ApJ, 596, 314}
\rf{Merritt, D., et al.  2002,
PRL, 88, 191301; 2009, ApJ, 693, L35}
\rf{Meyer, L. et al.  2008, ApJ, 688, L17; 2009, ApJ, in press}
\rf{Miralda-Escud{\' e} \& Gould 2000, ApJ, 545, 847}
\rf{Morris, M.\ 1993, 
ApJ, 408, 496}
\rf{Mouaward, N. et al.\ 2005, AN, 326, 83}
\rf{Murphy, B. et al. 1991, ApJ, 370, 60}
\rf{Najarro, F. et al.  1997, A\&A, 325, 700}
\rf{Nayakshin \& Cuadra 2005, A\&A, 437, 437}
\rf{Nucita, A. et al.  2007, PASP, 119, 349}
\rf{O'Leary \& Loeb 2008, MNRAS, 383, 86}
\rf{Olling \& Merrifield 2000, MNRAS, 311, 361}
\rf{Perets, H. B., et al. 2007, ApJ, 656, 709; 
2008a, in press;
2008b, IAUS, 246, 275}
\rf{Portegies-Zwart, S. et al.  2002, ApJ, 565, 265}
\rf{Quinlan, G. et al.  1995, ApJ, 440, 554}
\rf{Paumard, T. et al.  2006, ApJ, 643, 1011}
\rf{Reid, M. J., Brunthaler, A. 2004, ApJ, 616, 872}
\rf{Rubilar \& Eckart 2001, A\&A,  2001, 372, 95 }
\rf{Salim, S. \& Gould, A.  1999, ApJ, 523, 633}
\rf{Sanders, R.~H.\ 1992, 
Nature, 359, 131; 
1998, MNRAS, 294, 35}
\rf{Sesana, A. et al.  2007, MNRAS, 379, L45-L49;
2008, ApJ, 686, 432-447}
\rf{Sch\"odel, R. et al.\ 2002, Nature, 419, 694; 2003, ApJ, 596, 1015}
\rf{Tamblyn, P. et al.  1996, ApJ, 456, 206}
\rf{Ullio, P. et al.  2001, Phys. Rev. D, 64, 43504}
\rf{Weinberg, N. et al. 2005, ApJ, 622, 878}
\rf{Will, C. M. 2008, ApJ, 674, L25}
\rf{Young, P. 1980, ApJ, 242, 1232}
\rf{Yu, Q., Tremaine, S. 2003, ApJ, 599, 1129}
\rf{Yu, Q., et al. 2007, ApJ, 666, 919}
\rf{Yusef-Zadeh, F. et al.  2008, ApJ, 683, L147}
\rf{Zucker, S. et al.\ 2006, ApJLett, 639, L21}

\end{multicols}

\end{document}